\documentstyle[twocolumn,prl,aps,epsfig]{revtex}
\newcommand{\beq}{\begin{eqnarray}}
\newcommand{\eeq}{\end{eqnarray}}
\renewcommand{\vec}[1]{{\mathbf{#1}}}
\begin{document}
\draft

\title
{Can Short-Range Interactions Mediate a Bose Metal Phase in 2D?}
\author{Philip Phillips$^1$ and Denis Dalidovich$^2$}\vspace{.05in}

%
\address
{$^1$Loomis Laboratory of Physics\\
University of Illinois at Urbana-Champaign\\
1100 W.Green St., Urbana, IL, 61801-3080\\
$^2$National High Magnetic Field Laboratory\\
Florida State University, Talahasse, Florida, 31301}

%

\address{\mbox{ }}
\address{\parbox{14.5cm}{\rm \mbox{ }\mbox{ }
We show here based on a 1-loop scaling analysis
that short-range interactions are strongly irrelevant
perturbations near the insulator-superconductor (IST) quantum critical 
point.  The lack of any proof that short-range
interactions mediate physics which is present
only in strong coupling leads us to
conclude that short-range interactions
are strictly irrelevant near the IST
quantum critical point.  Hence, we argue that no new physics, such as the formation
of a uniform Bose metal phase can arise
from an interplay between on-site and nearest-neighbour interactions.  
}}
\address{\mbox{ }}
\address{\mbox{ }}

\maketitle

The standard model used to 
study\cite{doniach,mpa,wen,fisher,zwerg,chak2,ks,amb,wag2,otterlo,cha,frey,scal}  
the insulator-superconductor transition in thin
films is the commensurate Bose-Hubbard model or equivalently
the charging model for an array of Josephson junctions.  The Hamiltonian
for this model
\beq
\hat{H}=\frac{1}{2}\sum_{ij}\hat{n}_iV_{ij}\hat{n}_j-
J\sum_{\langle ij\rangle}\cos(\phi_i-\phi_j)
\eeq
consists of a charging term $V_{ij}$ and nearest-neighbour Josephson
coupling between grains possessing
a superconducting phase $\phi_i$.  The operator, $\hat{n}_i$ is the boson number operator
for site $i$.  In the on-site version of this model, $V_{ij}=\delta_{ij}V_0$,
where $V_0$ is the capacitance charging energy for each junction.  
The corresponding
free-energy functional for the on-site charging model lies
in the $O(3)$ universality class which possesses
a single quantum critical point\cite{doniach} signalling
the loss of phase coherence once $V_0/J$ exceeds a critical value.  

The recent experiments of Kapitulnik and co-workers\cite{kap} in which a
 metallic phase
has been observed to intervene between the superconductor and the eventual 
insulating phase suggests
that perhaps two phase transitions accompany the loss of phase coherence in
a 2D superconductor:  1) superconductor to Bose metal and 2) Bose metal
to insulator.  These experimental results
as well as earlier observations\cite{ephron,goldman,van} of a similar
metallic phase have stimulated a re-examination\cite{denis2,shim,kap2,dd} of the 
physics of phase-only
models.  In this context, Das and Donaich\cite{dd} have appended to the standard on-site
charging model a
nearest-neighbour charging term with amplitude $V_1$.
Concluding that $V_1$ is a relevant perturbation, 
they find that short-range interactions mediate a {\bf new} critical
point in which a Bose metal
phase obtains once the size of each grain is increased beyond a critical value
such that $V_1>V_0$.  The Bose metal
phase of Das and Doniach\cite{dd} is a uniform phase lacking both
phase and charge order.  Hence, this phase is translationally
and rotationally invariant.  This result is surprising because the critical
point in a Josephson junction array (JJA) is controlled by the standard
$\phi^4$ Wilson-Fisher critical point.  It is well-known
that short-range interactions are irrelevant near the Wilson-Fisher
critical point\cite{cardy}.  Moreover, no new critical point
is generated regardless of the magnitude of $V_1$.  Hence, the work
of Das and Doniach\cite{dd} stands in stark contrast to the standard view.
In addition, Fazio and Sch\"on\cite{fs} have analysed the nearest-neighbour charging
model as well and have shown that at $T=0$, no phase exists lacking
both charge and phase order.

Motivated  by the disagreement between
the standard picture\cite{cardy,fs} and the Das-Doniach result\cite{dd}, 
we take a closer look at the nearest-neighbour
charging model.  We show that as long as on-site Coulomb interactions
are present, that there is no
signature that screened interactions of any type are relevant
through 1-loop order.  Two conclusions are possible.  Either
the physics controlled by short-range interactions is
strictly a strong coupling problem with no weak-coupling
signature, or short-interactions are irrelevant 
at each order in perturbation theory in agreement with the standard view. 
The lack of any proof that short-range interactions flow to strong
coupling leads us to conclude that it is
unlikely that short-range interactions can mediate a new homogeneous
phases near the 
the quantum critical point associated with loss of phase coherence.
 
To establish this result, we write the partition function for the 
phase-only model
in the standard way as a path integral
\beq
Z=\int {\cal D}\phi e^{-S}
\eeq
where the statistical weight for each path,
\beq
S=\frac12\int d\tau\sum_k\frac{\dot\phi(\vec k)\dot\phi(-\vec k)}{V(\vec k)}
-\int d\tau\sum_{\langle ij\rangle} J_{ij} \cos(\phi_i-\phi_j),\nonumber\\
\eeq
defines the effective action for our problem.  Here,
$V(\vec k)$ is the Fourier transform of $V_{ij}$.  At the outset, we place no 
restriction on the range of $V_{ij}$.  To simplify this action, we first decouple
the charging term by introducing\cite{fisher} an auxilliary real gauge field,
 $A_0(\vec k)$, through
the identity
\beq
\exp \left[ -\frac{1}{2}\int d\tau\sum_{\vec k}{\dot \phi}(\vec k)
\frac{1}{V(\vec k)}
{\dot\phi}(-\vec k) \right]=  \nonumber\\
\int {\cal D} A_0 \exp \left\{ -\frac{1}{2}\int_
{\vec k,\omega}
\frac{A_0(\vec k,\omega)A_0(-\vec k,-\omega)}{e^{-2}V(\vec k)-1}\right.\nonumber\\
\left. -\frac{1}{2e^2}\int d\tau\sum_i
( {\dot\phi_i} -e A_0(i) )^2 \right\}.
\eeq
The coupling constant $e$ is a free parameter which will be determined
later.  The second step is to decouple 
the $\exp(\phi_i)$ terms in the standard way\cite{doniach} by introducing
the complex field $\psi_i(\tau)$, which will play the role of the order
parameter in an effective Landau-Ginzburg theory. The final
expression for the partition function
\beq
Z=\int{\cal D}\psi{\cal D} A_0 e^{-S}
\eeq
is obtained by integrating over the auxilliary fields.  The effective
action now takes the form,
\beq
S &=& \frac{1}{2}\int d^dx d\tau \left[
|(\partial_\tau-ieA_0)\psi|^2+|\nabla\psi|^2
+r|\psi|^2+\frac{u}{2}|\psi|^4\right] \nonumber\\
&+& \frac{1}{2}\int_{\vec k,\omega}\frac{A_0(\vec k,\omega)
A_0(-\vec k,-\omega)}{e^{-2}V(\vec k)-1}\nonumber\\
 &=& S_0+S_1.
\eeq

Consider first the case of long-range Coulomb interactions.  In this
case, the constant, $e$, plays the role of the electric charge, $e^\ast=2e$,
and $V(k)=(e^\ast)^2/k^\sigma$ where $\sigma=2$ for D=3 and $\sigma=1$ for
2D.  Consequently, the pure Coulomb part of the action reduces to 
\beq
S_1=\frac{1}{2}\int_{\vec k,\omega} k^{\sigma} 
A_0(\vec k,\omega) A_0(-\vec k,-\omega),
\eeq
which is the Fisher and Grinstein\cite{fisher} result.

What about short-range interactions?  We simplify to the case considered
by Das and Doniach\cite{dd} and truncate $V(\vec k)$ at
the nearest-neighbour level:
\beq
V(\vec k)=V_0+2V_1(\cos k_x+\cos k_y).
\eeq
It is crucial in our derivation that $V_0\ne 0$.  As has been considered 
previously, when $V_0=0$ but $V_1\ne 0$, the nature of the $T=0$ transition
changes fundamentally when compared with the $V_0\ne 0$ case.  In the former
case, that is, $V_0=0$ but $V_1\ne 0$, the $T=0$ transition is of the 
Berezinskii-
Kosterlitz-Thouless kind\cite{fs}.
In the long wavelength limit, $V(\vec k)=V_0+2V_1-V_1 k^2$, which
is convenient to write in the form, $e^2-V_1k^2$ where we have 
fixed the free parameter, $e^2=V_0+2V_1$.  Consequently, the pure gauge
 part of the action
simplifies to
\beq
S_1=-\frac{e^2}{2V_1}\int_{\vec k,\omega}
\frac{A_0(\vec k,\omega)A_0(-\vec k,-\omega)}{k^2}.
\eeq
Upon rescaling the gauge field, $A_0(\vec k,\omega)\rightarrow 
i\sqrt{\frac{V_1}{e^2}}A_0(\vec k,\omega)$,
we arrive
at the working form for the action,
\beq\label{action}
S&=& \frac{1}{2}\int d^dxd\tau \left[ |(\partial_\tau+gA_0)
\psi|^2+|\nabla\psi|^2
+r|\psi|^2+\frac{u}{2}|\psi|^4 \right] \nonumber\\
 &+&\frac{1}{2}\int_{\vec k,\omega}A_0(\vec k,\omega)\frac{1}{k^2}
 A_0(\vec k,\omega)
\eeq
where the constant $g=\sqrt{(V_1(V_0+2V_1)}=e\sqrt{V_1}$.  Clearly,
when $V_1=0$, $g=0$ and the rescaled fields, $A_0(\vec k,\omega)$, vanish
leading to the standard on-site charging model.  Hence, the relevance of 
short range
interactions can be deduced entirely from the scaling properties
of the coupling constant $g$.  

Performing the standard tree-level
rescaling with the rescaling parameter $b>1$, we find that
the momentum and frequency scale as $q'=qb$ and $\omega'=\omega b^z$,
with $z$ the dynamical exponent.  At the tree level, $z=1$ and the anomalous
dimension exponent vanishes, $\eta=0$.
Hence, the $\psi$ and $A_0$ fields scale as
\beq
&A_0=b^{\mu} A_{0}^{\prime}, \quad\mu=\frac{d+z-2}{2}&\nonumber\\
&\psi_0=b^{\lambda}\psi^{\prime}, \quad\lambda=\frac{d+2+z}{2}&.
\eeq
Combining these scaling relations with the rescaling of the momentum
and the frequency arising from the integrations in the action,
we arrive at our key result
\beq\label{g'}
g'=g\frac{b^{\mu+\lambda}}{b^{(3d+3z)/2}}=gb^{-(d+z)/2},
\eeq
namely that $g$ has a negative eigenvalue. Hence,
upon successive renormalization transformations,
the physics controlled by $g$ can have no effect on
the underlying quantum phase transition.  

\begin{figure}
\begin{center}
\epsfig{file=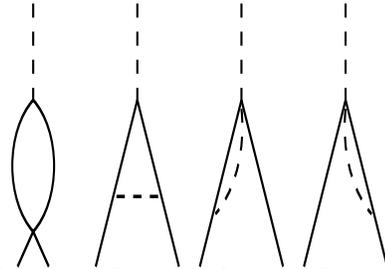, width=5.0cm, height=3.5cm}
\caption{Diagrams that contribute to the renormalization of $g$ through
1-loop order.  The dashed line represents the propagator for the
$A_0(\vec k,\omega)$ field.  The solid lines are given by the 
Gaussian propagator $(\omega^2+q^2+r)^{-1}$. }
\label{diag}
\end{center}
\end{figure}

Does the irrelevance of $g$ still persist beyond the tree-level?
To answer this question, we derive the scaling equations for $g$ through
1-loop order.  The relevant diagrams that contribute are shown in
Fig. (\ref{diag})\cite{ye}. 

We evaluate these diagrams using 
the standard frequency-momentum shell RG approach in which we integrate
out the fields $A_0(\omega,\vec k)$ and $\psi(\omega,\vec k)$ for momenta
and frequencies satisfying the constraint 
$\frac{\Lambda}{b}<\omega < \Lambda$,
and $\frac{\Lambda}{b} < k <\Lambda$ 
with the upper momentum and frequency cutoffs 
$\Lambda_\omega=\Lambda_k=\Lambda=1$.
Setting $b=e^\ell$, we
obtain
\beq
\frac{d g_\ell}{d\ell}=-\left( \frac{d+1}{2} \right) g_\ell-2A_\ell g_\ell 
u_\ell -B_\ell g^3_\ell
\label{dg}
\eeq
as the differential form for the scaling equation for $g$. The coefficients,
$A_\ell$ and $B_\ell$ are given by
\beq
A_\ell&=&\frac{2K_d}{(2\pi)^{d+1}}\left[\int_0^1 dq\frac{q^{d-1}}
{(q^2+1+r_{\ell})^2}\right.\nonumber\\
&&\left.+\int_0^1\frac{d\omega}{(1+\omega^2+r_{\ell})^2}\right]
\eeq
and
\beq
B_\ell&=&\frac{2K_d}{(2\pi)^{d+1}}\left[\int_0^1 dqq^{d+1}
\frac{1+2q^2+2r_{\ell}}
{(1+q^2+r_{\ell})^2}\right.
\nonumber\\
&&\left.+\int_0^1d\omega\frac{2+\omega^2+2r_{\ell}}
{(1+\omega^2+r_{\ell})^2}\right],
\eeq
where $K_d$ is the area of a $d$-dimensional unit sphere.
These coefficients are positive and depend on the scaling length $\ell$ through
the parameter $r_\ell$.  Hence, from the structure of the scaling equation,
for $g_\ell$, Eq. (\ref{dg}), we find that the $g=0$ fixed point is stable 
through one-loop order.  That is, there is no signature in weak coupling
that finite $g$ can drive a new critical point.  This
conclusion is consistent with the standard view that 
as long as the broken symmetry state is rotationally and translationally
invariant, the critical point is of the Wilson-Fisher type where it 
is well known that short-range
interactions cannot lead to a new critical
point. 

Because $g\propto \sqrt{V_1}$, the 1-loop scaling
equation for $g$ necessarily implies that
nearest-neighbour interactions are irrelevant and as a consequence
cannot change the phase diagram of the on-site charging
model, contrary to the claims of Das and Doniach\cite{dd}.  
Simply put, nearest neighbour interactions constitute an 
irrelevant perturbation in the phase-disordering transition
in complete agreement with the standard scaling arguments\cite{cardy}.
Because the critical point in the JJA model is of the Wilson-Fisher
type, $V_1$, regardless of its magnitude, cannot mediate a new critical
point separating phases that are translationally and rotationally invariant.
This rules out automatically a $V_1$-mediated uniform Bose metal phase.

However, short-range interactions
can mediate an inhomogenous charge-ordered
phase such as a supersolid\cite{frey,scal}.  In such instances, the effective
field theory reduces\cite{frey} to two coupled $O(3)$ vector models.  
Nonetheless, if the broken symmetry state is rotationally and translationally
invariant, the critical point is of the Wilson-Fisher type where short-range
interactions are strictly irrelevant. In light of this conclusion, the only
candidate for a Bose metal phase that remains 
is our recent proposal\cite{denis2} that in the standard
quantum disordered regime, a cancellation arises between
the exponentially long quasiparticle scattering time and the exponentially
small quasiparticle population, leading ultimately to a finite dc conductivity.

We thank E. Fradkin for
helpful comments.  This work was funded by the DMR98-12422 of the NSF research fund.

\end{document}